# Interfacial contribution to the dielectric response in semiconducting LaBiMn$_{4/3}$Co$_{2/3}$O$_6$

M. Filippi, B. Kundys, R. Ranjith, Asish K. Kundu, and W. Prellier

Laboratoire CRISMAT, CNRS UMR 6508, ENSICAEN,

6 Boulevard Maréchal Juin, 14050 Caen Cedex, France

Impedance measurements have been performed on a sintered polycrystalline sample of the perovskite LaBiMn$_{4/3}$Co$_{2/3}$O$_6$. Colossal dielectric permittivity often is measured in this class of semiconducting materials as a result of extrinsic factors. Our results show that a large offset in the capacitance, measured on a series of samples with different thickness, is due to the interfacial polarization. This contribution then can be removed from the data, creating a general procedure for dielectric measurements in semiconducting samples.

Colossal dielectric constants have been reported for several materials such as, for example, $CaCu_3Ti_4O_{12}$[1,2]. It has been proposed that such large values may result from artifacts owing to the charge injection at the electrode contacts[3,4]. This is particularly true for semiconducting samples when compared to insulators. This effect, known for decades, is always present in biological samples[5], and can increase the dielectric response towards colossal values (especially at low frequencies).

In the particular case of polycrystalline semiconducting samples, the role of grain boundaries also may contribute to anomalously large values of the permittivity[6,7]. As a result, it is important to extract the intrinsic values of the dielectric constant for a given material and to distinguish the various contributions. To obtain the response of semiconducting samples, it has been proposed previously to either perform an equivalent circuit analysis where the role of the depletion interfacial layers is represented by an element (A resistance in parallel with a capacitor (RC) or just a capacitor (C)) in series with the sample impedance[9-11], or avoid the formation of such layers by inserting a thin insulating material between the sample and the electrode[10,11]. Disagreement exists on this topic, and the anomalous dielectric responses have been assigned to electrodes polarization[3] or grain boundary artifacts[6]. In addition, a colossal dielectric constant with a large magneto-capacitance value is observed frequently in polycrystalline semiconducting samples with different cations responsible for their magnetic and dielectric responses[12].

The perovskite $LaBiMn_{4/3}Co_{2/3}O_6$ is a semiconducting samples with a large dielectric response. This material is also attractive because the Co for Mn substitution enhances the ferromagnetic transition temperature, leading to a $T_c$~130K, higher than in the other Bi-based perovskites[13].

In this letter, we have investigated the complex impedance (Z=Z'+iZ") of a series of $LaBiMn_{4/3}Co_{2/3}O_6$ sintered polycrystalline samples with different thicknesses, as a function of temperature and frequency. By measuring the thickness-dependence of the capacitance, we found that the low temperature data are not affected by artifacts and the material exhibits a frequency independent dielectric constant ε~40. At higher temperatures (T>120K), when the resistivity decreases, an interface effect appears, leading to a large offset in the capacitance versus the inverse of thickness plot. We conclude that the interfacial effect can be subtracted with this approach, opening the route to more careful measurements of dielectric constants in semiconducting samples.

The well sintered polycrystalline samples of $LaBiMn_{4/3}Co_{2/3}O_6$ were prepared by a conventional sol-gel method, and the complete synthesis and characterization details are reported elsewhere[13]. From a rectangular bar of material five samples were cut with thickness ranging between 0.46 and 2.05mm. Contacts were made using both silver paste and indium. The impedance measurements were carried out with a PPMS Quantum Design cryostat using an Agilent 4248A RLC bridge.

Figure 1 illustrates the as-measured dielectric permittivity and loss as a function of the temperature. At low temperature ε remains near 40 until it exhibits an initial sharp upturn (accompanied by a loss peak) around 120K, reaching values higher than 1000. Then, a second increase (indicated in the figure) occurs at higher temperatures, where a value as high as $10^4$ is reached. It should be noted that the curves show large frequency dispersion. All these features are commonly observed in ceramic semiconducting samples[3,4,10,11].

At low temperature (10K) the dc resistivity of the sample exceeds the maximum value that can be measured ($\rho > 10^7$ Ωcm) with the instrument used. Here the sample behaves as a pure capacitor (i.e. the current-voltage phaseshift is almost -90 deg). Upon increasing the temperature, a second contribution appears, which results in a low frequency arc in the Nyquist plot. Figure 2(a) summarizes the Z" vs Z' Nyquist plot at T=120K (where we observe the step in the permittivity) for different AC voltage amplitude between 100mV and 2V. The spectrum shows a low frequency dominating arc (right side of the fig.2a), with strongly non ohmic response to the AC voltage. The second arc at higher frequency is not sensitive to the voltage amplitude, and likely represents the intrinsic sample response.

In order to discern the origin of the observed features and obtain more quantitative information,

the capacitance was measured as a function of the inverse of the thickness 1/D. If a true bulk (where bulk means "non interfacial") response is detected, the capacitance should depend linearly on 1/D with a slope equal to $\varepsilon_0 \varepsilon A$ (where $\varepsilon_0$ is the permittivity of the vacuum, $\varepsilon$ is the permittivity of the material and A the area) and extrapolate to 1/D=0 (e.g. for infinite thickness).

In figure 3(a) we report the measured capacitance of the five samples as a function of the inverse of the thickness (1/D), at four frequencies between 1KHz and 300KHz. Data with a linear fit are shown for T=10K. The capacitance at each frequency is linear in reciprocal electrode spacing and extrapolates to zero. The obtained slope is as expected (i.e. with doubling 1/D the capacitance doubles), confirming that the intrinsic bulk sample response is measured.

In figure 3(b) a similar analysis is made at T=120K. The capacitance at 1KHz lies in the first non-ohmic arc of figure 2. It shows a linear dependence with thickness, and extrapolates through zero (within the error) with the expected slope. Hence the low frequency arc in fig.2 cannot be due to an interfacial effect, because the sample capacitance depends on the electrode spacing, while the polarization contribution does not. A calculation of the permittivity at this frequency (1KHz) gives an artificial value as high as 2000, and explains the upturn of "apparent" permittivity at 1KHz that starts at 100K (see fig. 1(a)). This Maxwell-Wagner like effect, with non-ohmic response, can be due to grain boundaries, as already proposed[6,7]. The present data seem to exclude an interfacial origin.

The C versus 1/D dependence is also linear at higher frequencies. The contribution from grain boundaries decreases with frequency, and at 10KHz and 100KHz persists (these frequencies lie in the region between the two arcs of fig. 2). Consequently, the capacitance depends on frequency. Finally for 300KHz we detect the true sample response, which yields a value $\varepsilon \cong 85$.

Figure 3(c) illustrates the C versus 1/D plot at 200K. Here, dc resistivity is low ($10^3$ $\Omega$cm) and a significant contribution from polarization of the electrodes appears. Although low frequency data are noisy, at higher frequencies the linear dependence of capacitance on 1/D appears clearly.

At this temperature, we expect a Maxwell-Wagner contribution due to the interfaces, resulting from a capacitance in series with the sample. The overall response is a resistance in parallel with a capacitor for the sample, in series with the interfaces capacitance. This second contribution will be thickness independent, leading to a constant shift in the measured capacitance at a given frequency and temperature. Consequently, the lines in figure 3(c) no longer extrapolate to zero, and a large offset is observed. The measured offset is strongly frequency dependent, and becomes larger at low frequencies. Thus, the parasitic interfacial capacitance, which is physically in series with the sample, results in a frequency dependent shift. The calculated permittivity, after subtraction, is $\varepsilon \cong 900$ ( this remaining large value is due to the first step in figure1, which is not interfacial)

In figure 4 we report the C versus 1/D plot at 200K obtained with indium electrodes. Also in this case, the low frequency data show the thickness independent offset. The line at 100KHz and 300KHz extrapolate to zero, demonstrating that for these frequencies interfaces are not relevant. Also at 300K (shown as inset), the interfaces only provide a slight contribution to the high frequency capacitance. The large permittivity (around $10^4$) measured at T=300K most likely is due to the higher conductivity of the sample.

We stress the importance of our results because in all the previous works on this topic (i.e. origin of giant permittivity and artifacts coming from conductivity) certain models with numerous parameters were used to fit the data. Here, we have derived a model from experimental data, where each element has been identified on the basis of its variation with frequency, AC voltage and sample thickness. We demonstrate that it is possible to extract the contribution of the interfacial polarization using a set of measurements for different sample thickness.

In summary, we have studied the dielectric properties of sintered polycrystalline perovskite $LaBiMn_{4/3}Ni_{2/3}O_6$ samples as a function of the thickness in order to evaluate the contribution coming from the electrode interfaces. We are able to separate the contribution coming from dipoles, from grain boundaries and from electrodes interfaces. This result appears of a general interest because it provides an experimental procedure to directly evaluate the interfacial polarization. We also report a large dielectric permittivity for LaBiMn$_{4/3}$Co$_{2/3}$O$_6$ compound which, when combined with its remarkable magnetic properties, makes this material suitable for technological applications. Unless careful measurements can be performed[14], any possible claim of magnetoelectric coupling should be interpreted cautiously because of these artifacts.

This work was carried out in the frame of the the STREP MaCoMuFi (NMP3-CT-2006-033221) and the STREP CoMePhS (NMP4 CT-2005-517039) supported by the European Community and the CNRS. The authors would also like to acknowledge W.C. Sheets, Ch. Simon, and J.F. Scott for careful reading of the manuscript.

**Figure Captions**

**Figure 1**.Temperature dependence of dielectric permittivity and loss at certain frequencies for a LaBiMn$_{4/3}$Ni$_{2/3}$O$_3$ ceramic sample. Contacts were made with silver paste.

**Figure 2.** Impedance complex plane plot (Nyquist plot) measured at 120K for various AC voltage amplitude between 0.1 and 2V. Frequency was swept from 20Hz to 1MHz. Contacts were made with silver paste.

**Figure 3.** Sample capacitance as a function of the reciprocal thickness 1/D measured at 10K (a) 120K (b) and 200K (c). Data are shown for frequencies between 1kHz and 300kHz.

**Figure 4.** Sample capacitance as a function of the reciprocal thickness (1/D) recorded at 200K, for contacts made with indium. The same plot taken at T=300K is shown as inset.

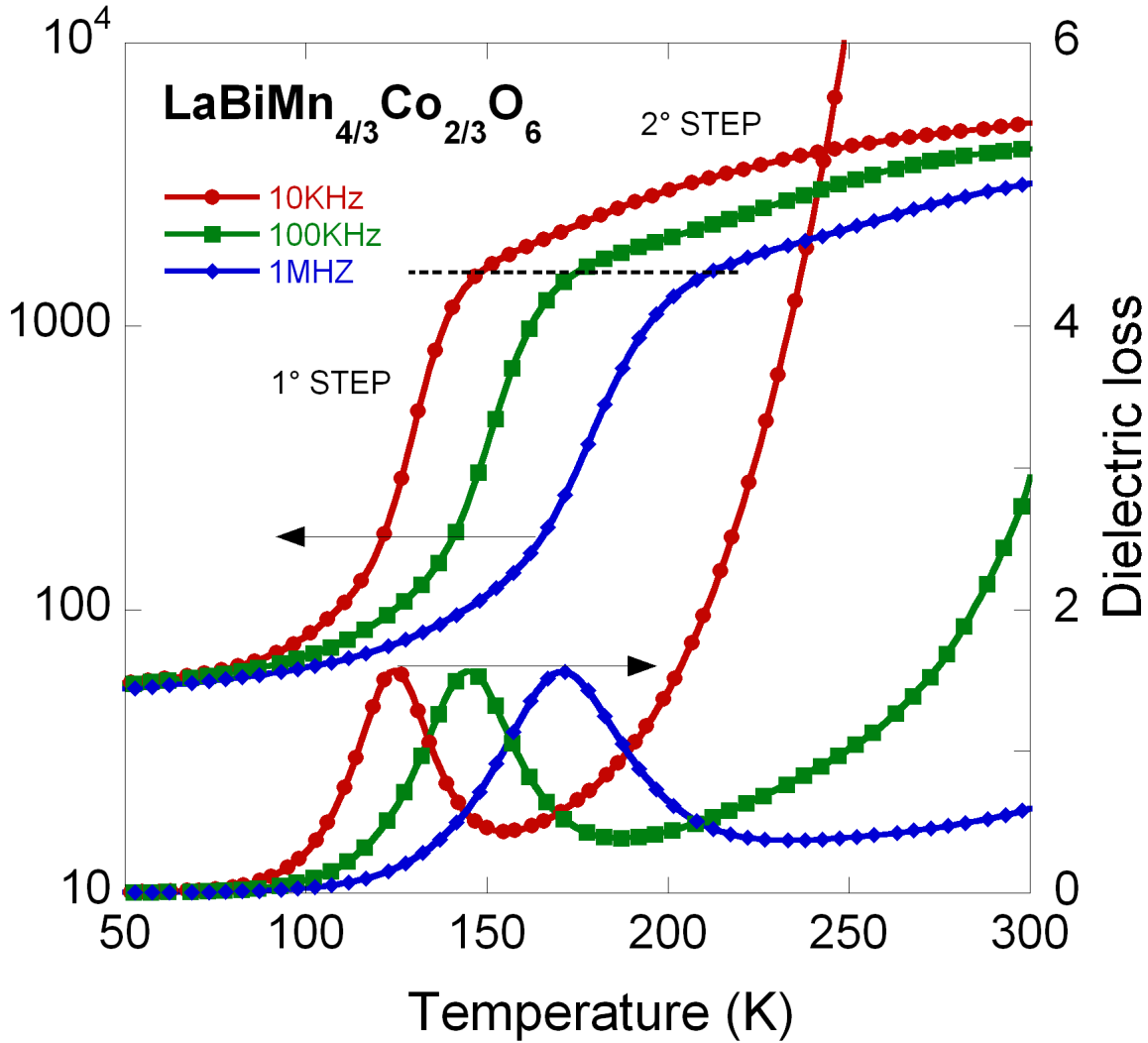

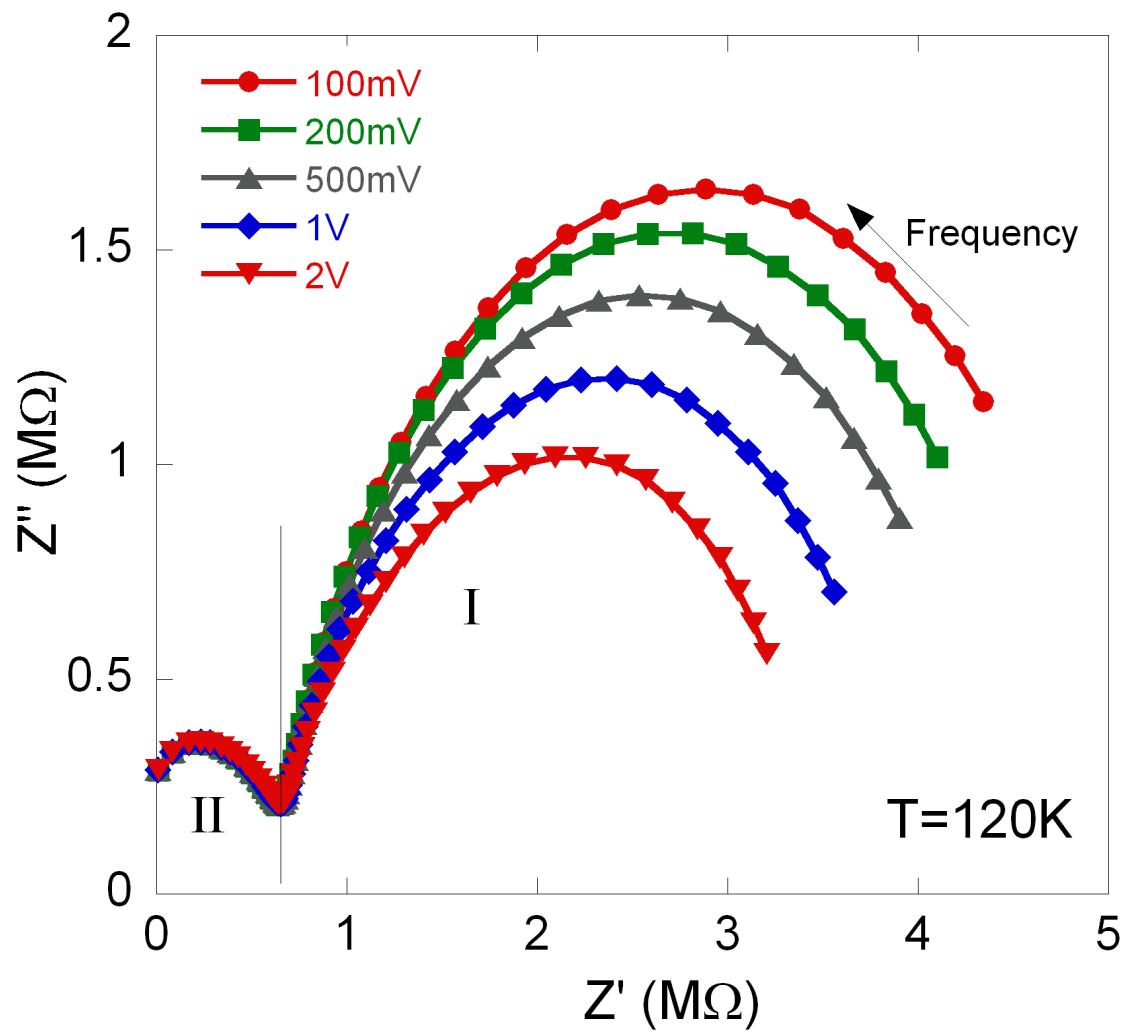

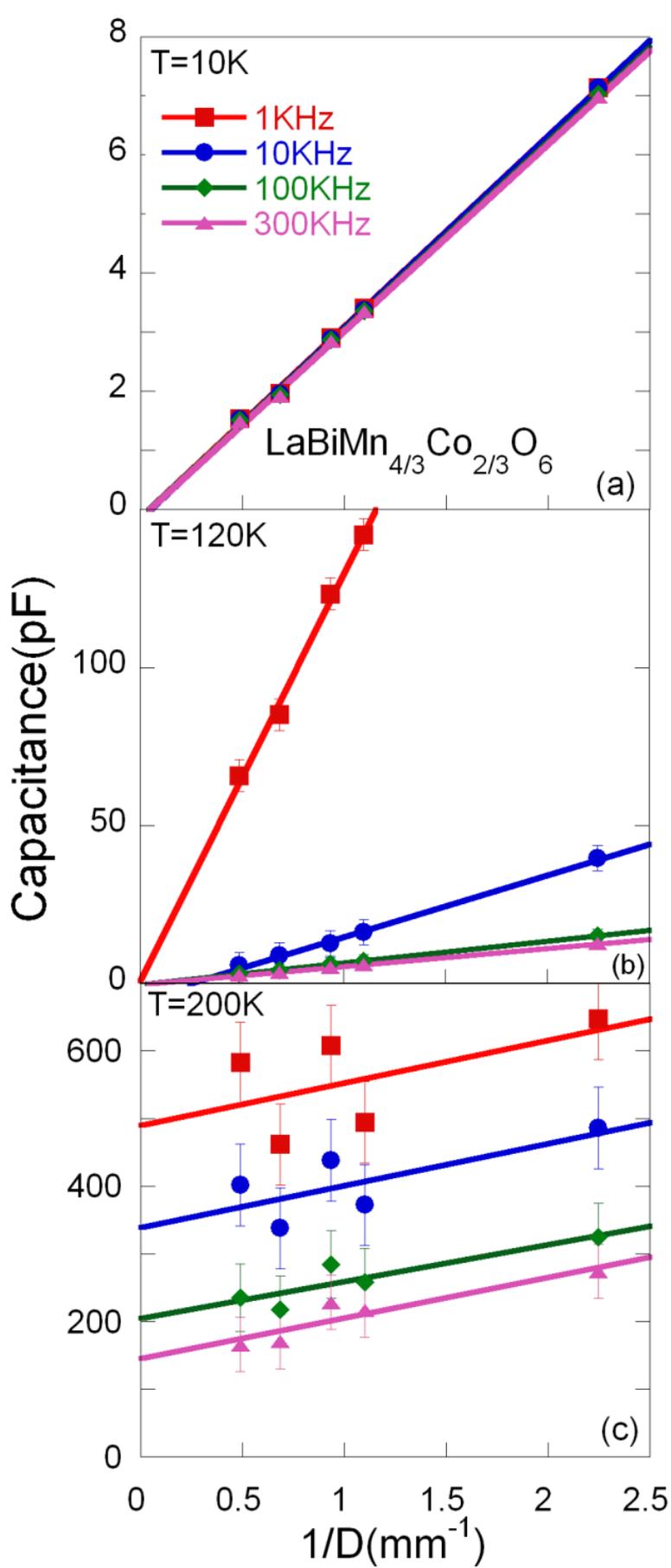

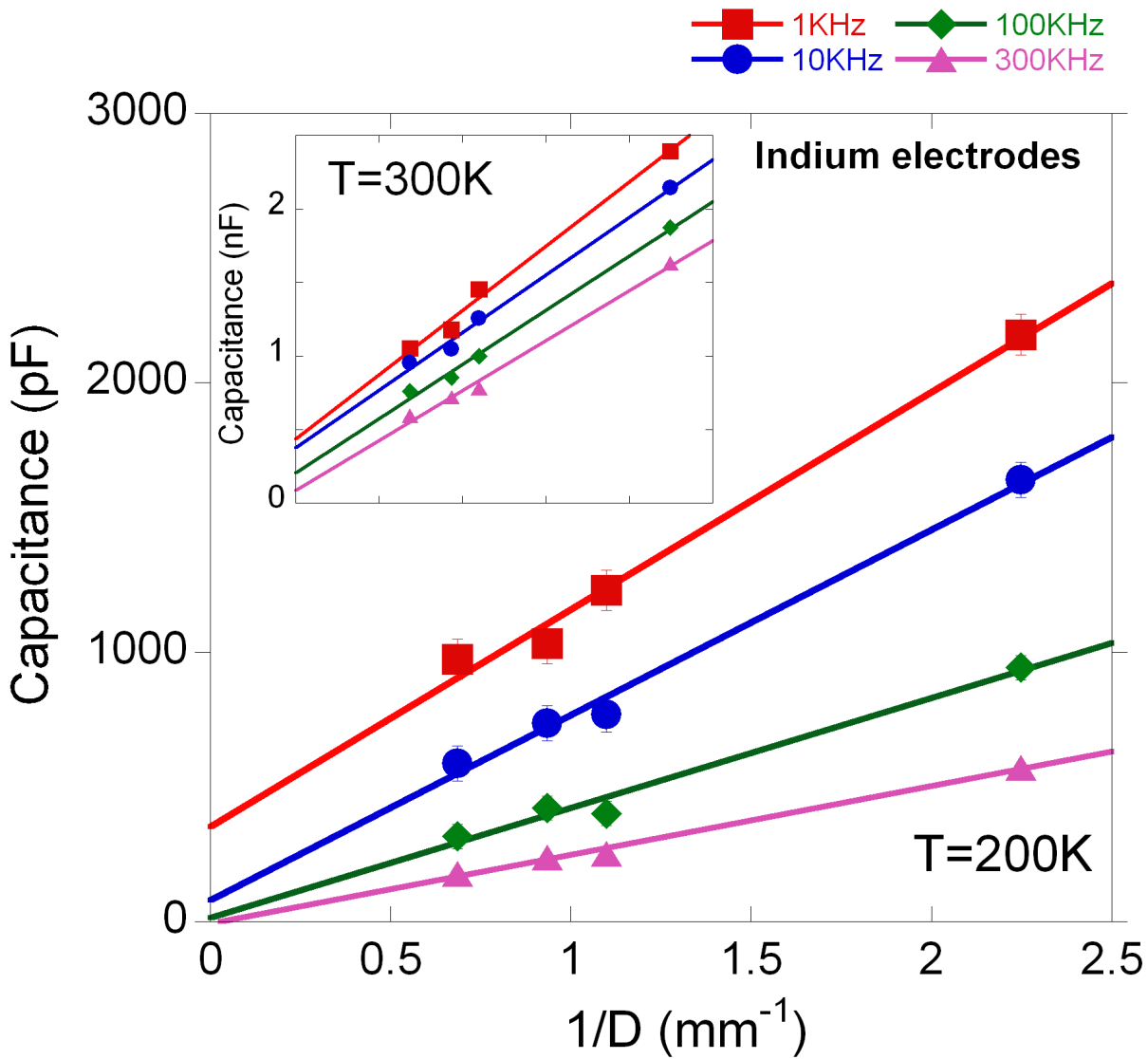